# Superconducting single-photon detector integrated in DBR with optical microconnector for MM or SM fiber

Maksim V. Shibalov[1], Ilya M. Asharchuk[1], Evgeniy O. Epifanov[3], Igor V. Trofimov[1], Alexander M. Mumlyakov[1], Nikita V. Minaev[3], Yuri V. Anufriev[1,2] and Michael A. Tarkhov[1,2]

[1]Institute of Nanotechnology of Microelectronics of the Russian Academy of Sciences
[2]National Research University "Moscow Power Engineering Institute"
[3]National Research Centre "Kurchatov Institute"

This paper presents the development of a superconducting nanowire single-photon detector (SNSPD) integrated into a distributed Bragg reflector (DBR) with a design center wavelength of 830 nm and a width of 200 nm. This SNSPD is made of a superconducting niobium nitride (NbN) thin film that is produced using plasma-enhanced atomic layer deposition (PEALD). The DBR is made of 15 alternating layers of silicon nitride and silicon oxide that are produced through plasma-enhanced chemical vapor deposition (PECVD). The reflection efficiency of the mirror is 90% at a wavelength of 830 nm. For sufficient optical coupling, an optical micro-connector optimized for multimode or single-mode optical fibers with a diameter of 128 μm was formed using two-photon polymerization techniques. The niobium nitride film was deposited onto the DBR surface in-situ in two separate reactors connected by a vacuum transfer. The in-situ technique of deposition of a superconducting niobium nitride film and a distributed Bragg reflector has allowed achieving a detection efficiency of 90% at a wavelength of 830 nm and a dark count rate of 10 $s^{-1}$ at a temperature of 2.5 K. Additionally, the detector jitter was 50 ps.

## I. INTRODUCTION

Recent studies [1-4] have demonstrated the potential and exceptional quality of utilizing ultrathin superconducting niobium nitride films that are obtained through atomic layer deposition. These films have been used successfully for creating superconducting single-photon detectors and high-Q resonators [5-7]. One of the main advantages of this technology is the ability to produce thin films with precise control of thickness and specified parameters on wafers with a diameter of 4 inches or more, ensuring high uniformity [8, 9]. Recent research related to the uniformity of thin films of niobium nitride obtained by atomic layer deposition has revealed that deviations in the uniformity of the superconducting transition temperature and the critical current can be achieved within 1% and 12%, respectively, using 2-inch wafers [10]. Therefore, atomic layer deposition technology is constantly developing for a scalable manufacturing cycle of superconducting devices with various functional purposes [11-13]. The use of a distributed Bragg reflector based on dielectric layers with different refractive indices [14-16] is the most favorable way to increase the quantum efficiency of SNSPDs. To ensure excellent deposition uniformity and layer roughness throughout the silicon substrate [17, 18], plasma chemical techniques for forming thin dielectric films for such reflectors are used. Combining these techniques with atomic layer deposition provides a significant advantage in producing large and efficient arrays of single-photon detectors on substrates with a length of 100 mm.

In this study, distributed Bragg reflector (DBR) detectors consisting of 15 optical pairs were used to enhance the reflection and absorption of the detector sensing element. It is noteworthy that the dielectric DBR layers and the superconducting ultrathin niobium nitride film were deposited in-situ in two separate reactors, which were connected through a vacuum transfer. This ensured the lack of absorbent substance on the substrate surface that could interfere with the growth of the NbN film through the PEALD technique. As a result, the air oxidation of the upper layer of silicon nitride was prevented [19], which could have had an adverse effect on the growth of a niobium nitride film with a thickness of several nanometers.

To ensure precise coupling of the optical fiber and detector's sensing element, we utilized the two-photon polymerization technique which enables the creation of localized optical fiber microconnectors that are optimized for either MMF or SMF [20]. These microconnectors effectively secure the optical fiber directly above the detector sensing element, resulting in an exceptionally efficient coupling of the SNSPD sensing element and the fiber core close to 1 due to geometric overlap. Thanks to the micro-connector's dimensions of approximately 1000 μm·650 μm·200 μm (H·L·W), the process of aligning and leading the fiber using micromanipulators was simple. The manufactured detectors delivered a system efficiency of approximately 90% at a wavelength of 830 nm. This wavelength is used in biological tissue spectroscopy, single quantum emitters, and other applications [21, 22].

## II. FABRICATION

The substrate selected was a *P*-type silicon wafer <100> with a diameter of 4 inches. Following the initial washing, seven pairs of silicon nitride ($SiN_x$) and silicon oxide ($SiO_x$) were created on the wafer through the employment of the PECVD technique within an Oxford Instruments Plasmalab 100 reactor, with the layers being alternated pairwise. A layer of silicon nitride was deposited onto the last silicon oxide layer. The silicon nitride layers were deposited in a gas mixture of monosilane, ammonia, and nitrogen ($SiH_4/NH_3/N_2$) at a pressure of 1,250 mTorr and an RF source power of 30 W. Following the deposition of the silicon nitride layer, the process chamber was purged and pumped out for five minutes. A layer of silicon oxide was then deposited utilizing a gas mixture of tetraethoxysilane and oxygen ($TEOS/O_2$) at a pressure of 400 mTorr and an LF source power of 70 W. After deposition of silicon oxide, the chamber was purged and pumped out, which lasted for five minutes. This particular sequence was then repeated in succession until the requisite number of pairs of DBR layers was achieved. A comprehensive procedure for the DBR fabrication and measurement is presented in the research [23]. After the deposition of the DBR dielectric layers, the wafer was transferred to an Oxford Instruments FlexAL ALD system, maintaining the vacuum cycle throughout the process.

The fabrication of the niobium nitride film was achieved through the employment of plasma-enhanced atomic layer deposition. The deposition process is described in detail in our previous work [24]. The number of deposition cycles was 93. The meander-shaped sensing element was fabricated by utilizing electron beam lithography followed by plasma-chemical etching. The etching process of niobium nitride through a resistive mask was executed via the plasma-chemical technique with a gas mixture $SF_6/Ar$ at a gas flow ratio of 2:1. The contact pads were generated from aluminum utilizing the lift-off technique. Subsequently, the fabricated detectors were split into 5x5 mm chips using a sawing technique. Finally, a cleansing process employing water under pressure was carried out to eliminate any contaminants.

## III. RESULTS AND DISCUSSION

The meander dimensions were 7.5x7.5 μm with a width of 100 nm and a filling factor of 0.5. Figure 1(a) displays an SEM image of the sensing element, and the insert to Figure 1(a) provides a closer view of an area of the sensing element of feature dimensions. To demonstrate the distributed Bragg mirror and meander multi-layered structure, a lamella was created and TEM microscopy was carried out, which is displayed in Figure 1(b). The TEM image revealed that the niobium nitride film's thickness was 5.6 nm. Figure 1(c) exhibits a Comsol Multiphysics simulated DBR wavelength dependence of reflectance, which contains seven pairs of $Si_3N_4/SiO_2$ layers with a $Si_3N_4$ cap layer represented by the black line. The experimental wavelength dependence of reflectance in the 650-1000 nm range is illustrated by the red line. The measured reflection efficiency is approximately 90% at the wavelength of 830 nm.

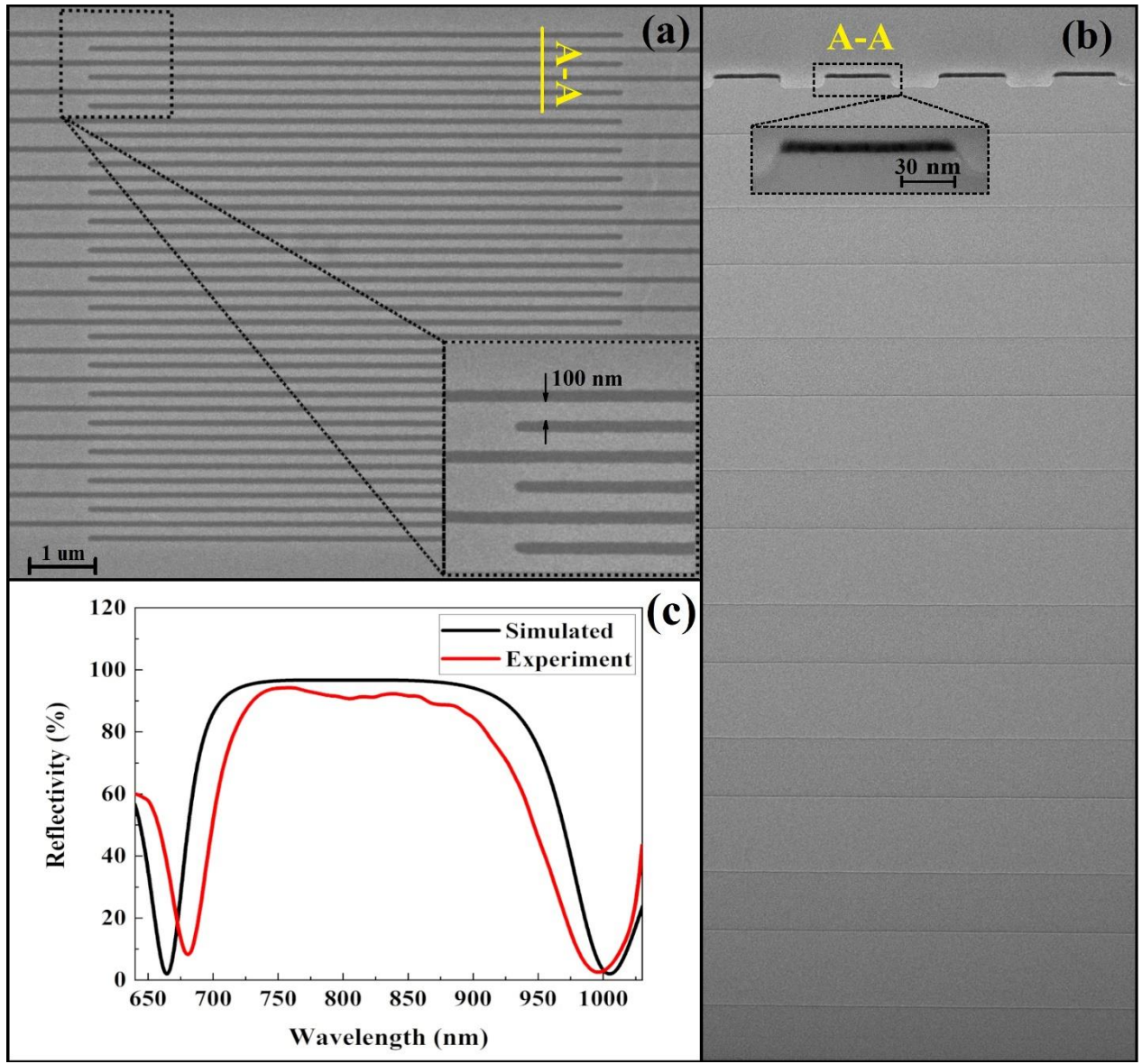

FIG. 1. SEM image of an SNSPD sensing element with a strip width of 100 nm (a). TEM image of 15 layers of DBR's and SNSPD sensing element (b). The inset in (b) shows an enlarged image of the sensing element strip. Calculated (black) and experimentally measured (red) wavelength dependence of DBR reflection (c).

Using the two-photon polymerization technique, an optical microconnector was produced directly above the detector sensing element. This process utilized a two-photon micro-stereolithography unit equipped with the ABL1000 submicron positioning system, manufactured by Aerotech, USA. The radiation source employed was the second harmonic of the TEMA-100 femtosecond laser with a wavelength of 525 nm, a pulse duration of 200 fs, and a frequency of 70 MHz, manufactured by Avesta-Proekt, Russia.

The process of creating an optical microconnector involves four distinct stages. The first stage involves modifying the DBR surface using laser radiation in a nonlinear mode through a process of laser ablation. In the second stage, a spacer layer of a photopolymer composition (FPC) is installed on the sample, followed by another stage implying the femtosecond polymerization technique, which involves polymerization in the volume of a photopolymer composition transparent to the operating wavelength of radiation. As a result of the two-photon absorption, a polymerization begins which leads to the hardening of a volumetric area within the FPC volume. The installation software represents a 3D model of an optical microconnector in the form of layers, which is subsequently printed layer-by-layer onto a substrate surface that has been previously modified. At the last stage, the sample is washed with isopropyl alcohol and a small amount of chloroform to remove any excess FPC. After being washed several times, the finished optical microconnector is air-dried.

Figures 2(a), (b), and (c) display images of the microconnector formed on the detector in different planes captured through SEM microscopy. Figure 2(a) also features an insert showcasing the surface modification aimed at enhancing the adhesion of the polymer material used in optical microconnector manufacturing. The microconnector's design considers the mechanical actions of the fiber during the optical coupling of the MMF or SMF with the SNSPD sensing element.

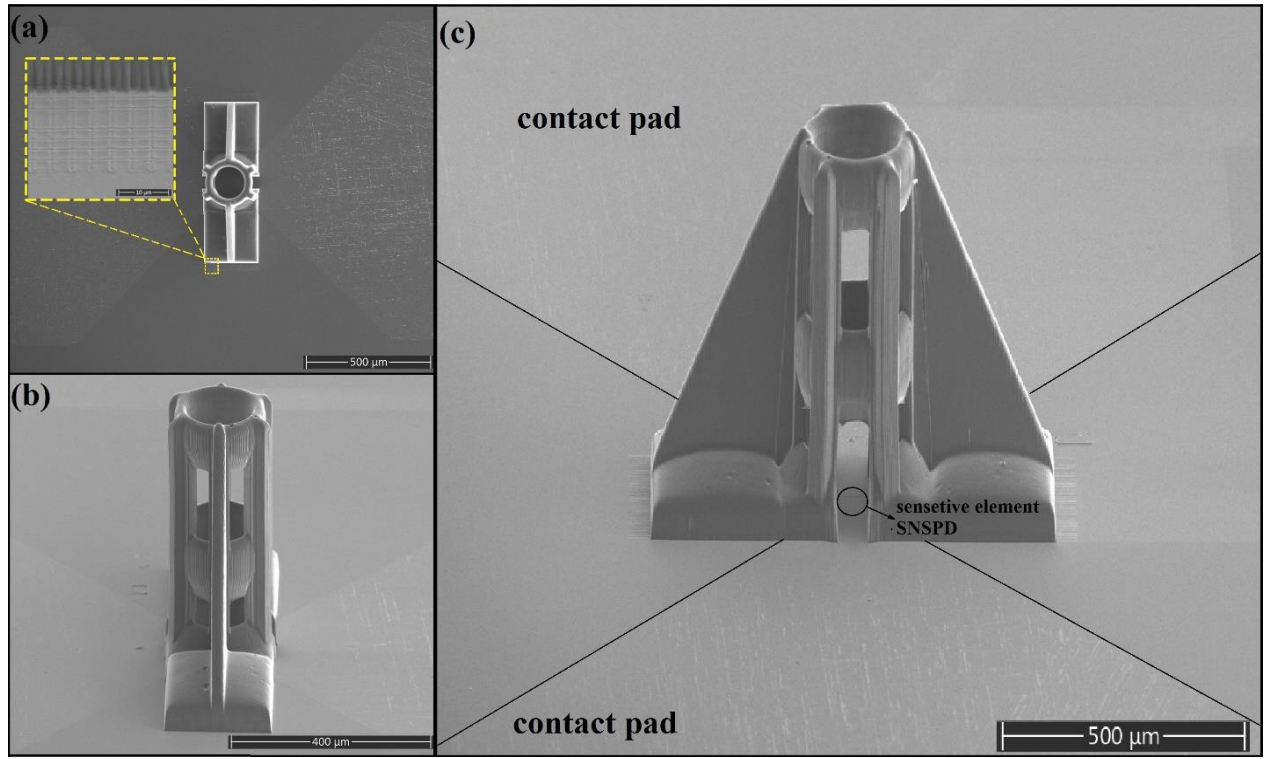

FIG. 2. SEM images of the optical micro-connector at different imaging angles (a), (b), (c). The inset to Fig. 1 (a) shows a surface modification to improve micro-connector adhesion.

In order to establish electrical contact, the sample was ultrasonically welded onto a microstrip line that was coupled with an SMA connector. Figure 3(a) shows the setup that was used to connect the fiber to a SNSPD mounted in a cooled holder. To achieve this, the holder with the sample was securely installed on a table that could move along the X and Y axes using micromanipulators. A digital camera was installed opposite the sample holder to enable observation of the optical microconnector. Directly above the microconnector, an optical SMF was mounted on a micromanipulator. Thanks to the micromanipulator, the optical fiber was capable of moving along the Z axis. The fiber's other end was fixed in an optical circuit that enabled the introduction of modulated radiation from a CW laser with a wavelength of 625 nm. A lock-in amplifier with bolometric response was used to record a signal, and the quality of the optical coupling was evaluated at the modulation frequency of the CW laser. The sample holder's SMA connector was linked to the DEXINMAG DXA-002 DSP lock-in amplifier via a coaxial cable. As a result of micromanipulators moving along the X and Y axes, the sample and the connector are coupled with the optical fiber. With the aid of a micromanipulator moving along the Z axis, the optical fiber is inserted into the connector. The step-by-step insertion process can be seen in figure 3(b), (c), (d), and (e). The optimal depth of fiber insertion inside the optical microconnector was determined based on the maximum value of the bolometric response, which was controlled using a lock-in amplifier. According to the signal obtained from the lock-in amplifier, it is possible to determine the complete contact between the fiber end and the SNSPD sensing element. The schematic representation of this process is given in Figure 3(a). The optical fiber within the microconnector was secured using ultraviolet glue. After that, the holder containing the sample and the optical fiber that is fixed in the microconnector is subsequently mounted within the Gifford-McMahon closed-cycle cryostat.

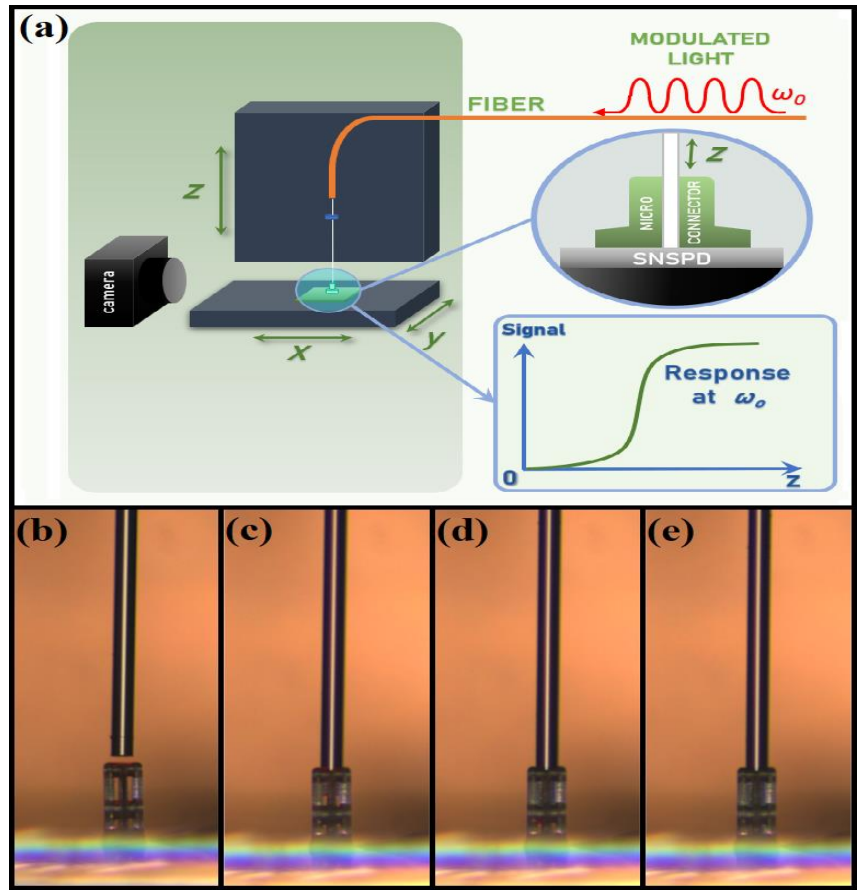

FIG. 3. Schematic representation of the process of aligning an optical fiber and a micro-connector (a). The step-by-step process of inserting an optical fiber into the micro connector (b), (c), (d), (e).

The sensing element in the cryostat was illuminated at a temperature of 2.5 K, with radiation injected through a single-mode optical fiber. The sensitive element was illuminated using of Thorlabs LPS-FC FW lasers emitting light at wavelength of 830 nm. The output power of the lasers was monitored using the Thorlabs PM200 optical power and energy meter. Attenuators of various ratings were utilized to reduce the laser radiation power. To achieve separation of the detector direct current bias from the detector RF output pulses, a Mini-Circuits ZFBT-282-1.5A+ bias tee was employed. To amplify the output HF pulses, a stage of Mini-Circuits ZX60-6013E-S+ RF low-noise amplifiers was utilized.

The temporal instability of the leading edge, or jitter, was measured with a pulsed fiber laser that operates at 1550 nm and has a pulse duration of around 80 fs and a frequency of 50 MHz from Avesta-Proekt. The laser radiation output was divided into two optical channels at a ratio of 50:50. The first channel led to an SNSPD, and the second led to a high-speed photodetector. The electrical output signals from both detectors were sent to a Becker and Hickl correlator board (SPC-150-NXX). This helped us determine detector jitter through statistical signal processing.

Figure 4(a) displays the photon count rate curve for a wavelength of 830 nm, revealing a plateau caused by saturation in the detection efficiency. Our system's quantum detection efficiency was approximately 90%, as determined by the DBR reflection efficiency. The dark count curve experiences an exponential increase with an increase in bias current. In Figure 4(b), a typical output pulse from the detector is depicted. The insert to Figure 4(b) presents a histogram of the change in the time of pulses coming from the detector over time. The obtained data were processed using the Gaussian function. The value of Δt, defined as FWHM, represents the timing jitter, with a value of 50 ps.

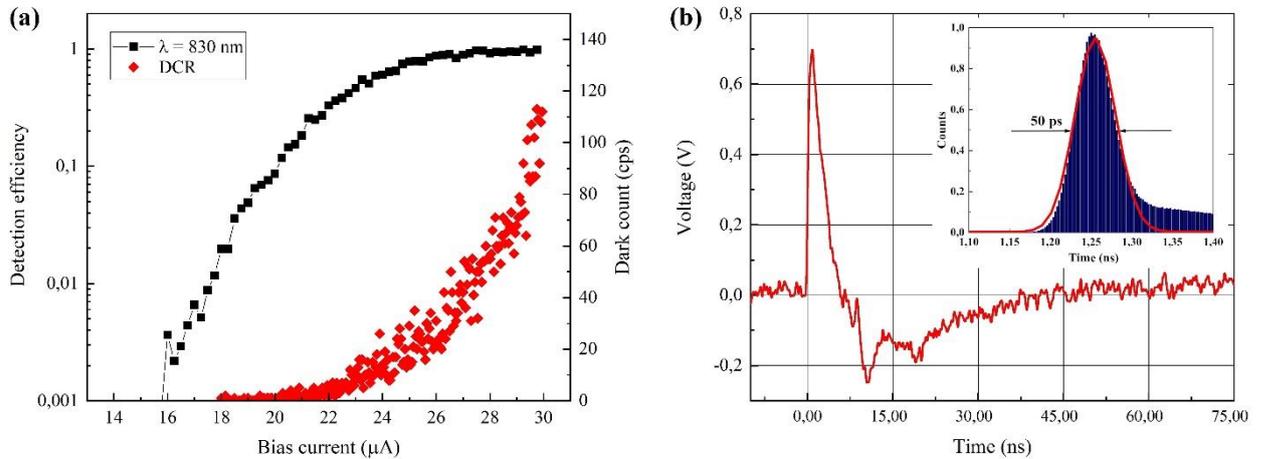

FIG. 4. Dependence of the count rate on the bias current for illumination with photons at a wavelength of 830 nm and dark count (a). Example of an output pulse from the SNSPD (b). Inset: timing jitter histogram for SNSPD at 1550 nm wavelength. The data have been fitted with a Gaussian function.

## IV. CONCLUSION

In this study, we have demonstrated SNSPD based on an NbN thin film produced using the PEALD technique integrated into a high-efficiency resonator with enhanced optical coupling using an optical microconnector. Our system has achieved an SDE of approximately 90% at a wavelength of 830 nm. The use of optical microconnectors has significantly increased the efficiency of optical coupling between the SNSPD sensing element and the optical fiber. Furthermore, using optical microconnectors has simplified optical coupling when dealing with arrays of multiple SNSPDs, as the connectors are small and allow for independent optical channels to each SNSPD in a small area. Our upcoming research will delve deeper into the homogeneity of NbN thin films obtained through the PEALD technique while analyzing the behavior of superconducting transition temperature, critical current density, and quantum detection efficiency on a 100 mm silicon substrate. Moreover, we aim to develop comparable detectors on a single chip, featuring up to 16 independent optical inputs.

## ACKNOWLEDGMENTS

The authors would like to thank E. Pershina and E. Eganova for their assistance in TEM imaging. The authors are grateful to Grant No. 122040800157-8 in support of the Ministry of Science and Higher Education of the Russian Federation. Fabrication and characterization were carried out at large-scale facility complexes for heterogeneous integration technologies and silicon + carbon nanotechnologies.